\documentclass[prc,aps,showkeys,showpacs,nofootinbib,twocolumn]{revtex4}
\usepackage{graphicx}  % uncomment this if your want to insert figures

\begin{document}

%%%%%%%%%%%%%%%%%%%%%%%%%%%%%%%%%%%%%%%%%%%%%%%%%%%%%%%%
% The title, only the first letter capitalized; if you want to split it in
% two or more lines, put a \\ macro at each line break
% example:
%   \title{Title: first line\\ second line}
%
\title{Phase-space methods in nuclear reactions around the Fermi energy}

%%%%%%%%%%%%%%%%%%%%%%%%%%%%%%%%%%%%%%%%%%%%%%%%%%%%%%%%
% The author(s), separated by commas; do not put a
% comma before the last author, use instead the \and
% macro which produces a normal ``and'' in the
% caps/small caps context
%
\author{\underline{Denis Lacroix}\footnote{lacroix@lpccaen.in2p3.fr},
Dominique Durand, Gregory Lehaut, Olivier Lopez and Emmanuel Vient}

%%%%%%%%%%%%%%%%%%%%%%%%%%%%%%%%%%%%%%%%%%%%%%%%%%%%%%%%
%
%\organization{Laboratoire de Physique Corpusculaire, \\
\address{Laboratoire de Physique Corpusculaire, \\
ENSICAEN and Universit\'e de Caen,IN2P3-CNRS,\\
Blvd du Mar\'{e}chal Juin
14050 Caen, France}

\begin{abstract}
Some prescriptions  for in-medium complex particle production
in nuclear reactions are proposed. They have been implemented in
two models to simulate nucleon-nucleus  (nIPSE)
and nucleus-nucleus (HIPSE\footnote{The computer program of the model is available at \\
%{\bf
${\bf http://caeinfo.in2p3.fr/theorie/theory\_lacroix.html}$
%}
}) reactions around the Fermi energy \cite{Lac04,Lac05}.
Our work emphasizes the effect of randomness in cluster formation,
the importance
of the nucleonic Fermi motion as well as the role of conservation laws.
The key role of the phase-space  exploration before and after 
secondary decay is underlined. This is illustrated in the case of two debated issues: 
the memory loss of the entrance channel 
in central collisions and the $(N,Z)$ partitions after the
pre-equilibrium stage.

\end{abstract}

\maketitle

%%%%%%%%%%%%%%%%%%%%%%%%%%%%%%%%%%%%%%%%%%%%%%%%%%%%%%%%
% Write the text starting from here and using the usual
% LaTeX commands.
%

In nuclear reactions around the Fermi energy,  nuclei
can break into several pieces of various sizes: the so-called multifragmentation process\cite{Dur01}. A
striking experimental feature  is the large number of accessible charge and energy partitions. In order to understand 
the statistical aspects of the explored phase-space, several
scenarii have been proposed.
However, in view of  the complexity of nuclear reactions due for instance to  impact
parameter mixing, pre-equilibrium emission
and thermal decay, it is hard to trace back the process of
cluster formation. This issue remains a highly debated
question. During the past three years, we have try to answer the following question: "What are the minimal hypothesis required
to reproduce the characteristics of nuclear fragmentation?". Based on a detailed and a successfull 
comparison with experimental data, we have ended up  
with surprising
conclusions. Simple prescriptions  have
been found for the formation and the emission of complex particles \cite{Lac04,Lac05,Lac05-2}.  
First, these rules are  introduced and illustrated. In a second part, we discuss two
aspects of current interest related to the phase-space exploration after pre-equilibrium stage.

\section{Rules for the formation and the emission of clusters}

%\begin{figure}
%\includegraphics{figurename}
%\caption{Your caption here}
%\label{fig01} % optional figure label, must be unique
%\end{figure}

\begin{figure}[tbph]
\begin{center}
\includegraphics[height=8.cm,angle=-90.]{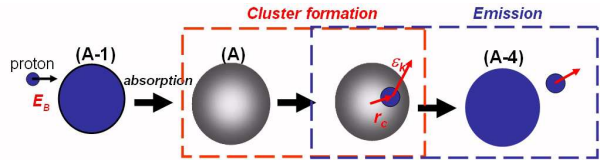}
\end{center}
\caption{ Schematic representation of a three step nucleon-induced
reaction. A nucleon with  energy close to the Fermi energy is
first absorbed by a nucleus. Then two steps are identified for pre-equilibrium emission: the
formation of the cluster and its emission.}
\label{fig:clus1}
\end{figure}

\begin{figure}[t]
\begin{center}
\includegraphics[height=9.cm,angle=-90.]{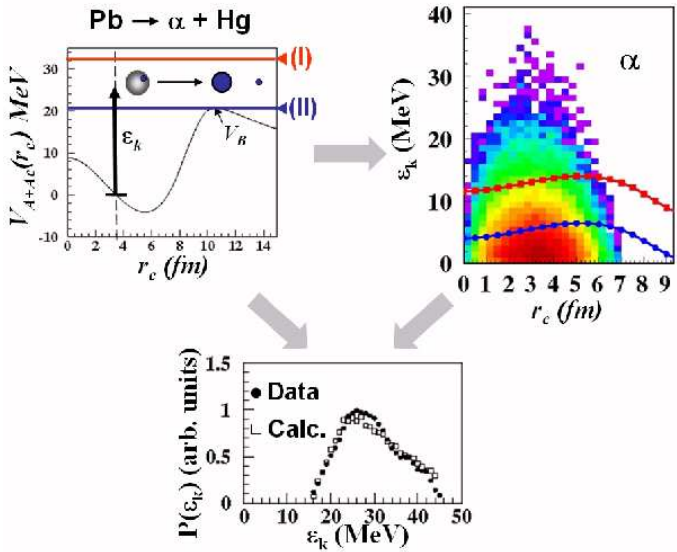}
\end{center}
\caption{ { Top-left: Two-body potential between the $\alpha$ and the
emitter. Above the line (I), the
cluster cannot be emitted. This upper limit is directly given by
the energy balance of the reaction.
Below the line (II), the cluster cannot overcome the barrier (since here,
quantum tunnelling is not considered). In between the two lines, there is
a small
"escape window" for the emission of the cluster.
Top-right: Total available phase-space of the cluster. This latter  
is significantly
reduced due to the energy constraint. The two curves correspond
respectively to the lower and
upper limit in the kinetic energy.
Bottom: Calculated kinetic energy distribution (open squares)
of the $\alpha$ particle obtained by propagating each
configuration in the "escape window" up to infinity. The
calculated spectrum is compared
with the experimental data (black circles).
}%end it
}
\label{fig:clus2}
\end{figure}

In this section, rules for the cluster formation and the production of
fragmentation partitions are defined. These rules have been extensively discussed in ref. \cite{Lac05-2} and
we only give here a short summary.
Let us  consider a proton colliding a heavy
target with an energy $E_B$ close to the Fermi energy.
A schematic  three steps
scenario is considered (see figure \ref{fig:clus1}).
First, the incident nucleon is absorbed. The second
step corresponds to the in-medium formation of the cluster.
We assume that clusters are made of  nucleons randomly picked 
inside a Thomas-Fermi distribution. Cluster
properties, position $r_c$ and the kinetic energy $\varepsilon_k$
are calculated on the basis of the characteristics of the nucleons in phase-space. Using a Monte-Carlo sampling,
an ensemble of configurations (called {\it
"total accessible phase-space"}) is obtained. As an illustrative example,
the phase-space of an $\alpha $ particle is given in top right of figure \ref{fig:clus2}.
The last step corresponds to cluster emission.
All configurations obtained in the second step  will not
necessarily lead to the emission of a cluster.
Two constraints on the emission can be identified. The first one,
which is independent of the entrance
channel, is due to the mutual interaction between the cluster and
the other nucleons. Given an interaction potential between the cluster
and the emitter, denoted by $V_{A+A_c}(r_c) $,
the emitted particle must overcome the barrier, denoted by $V_B$. This gives
a lower limit on the cluster kinetic energy.
The second constraint is directly dependent on the reaction type and is
due to the energy
balance. The constraint of a positive excitation energy for the configuration
leads to
\begin{eqnarray}
E_{B} - Q - V_{A+A_c}(r_c) \geq \varepsilon_k(r_c)
\end{eqnarray}      
where $Q$ is the Q-value of the configuration. Therefore, only a fraction of the
total phase-space accessible for the cluster will indeed lead to emission in the continuum.
This fraction corresponds to the {\it "explored phase-space"}.
These two constraints are shown in Fig. \ref{fig:clus2} (top left)
for a proton-induced reaction at $E_B = 39$ MeV.
There, an $\alpha$ particle can only be emitted in a small
interval of kinetic energy (called "escape window"
in the following) leading to a significant
restriction in phase-space. According to the
energy constraint, all configurations
between the two lines lead to the emission of an $\alpha$ particle (Figure \ref{fig:clus2} (top-right)).

If we now apply the rules and propagate classically all explored configuration with the interaction potential,
we obtain a distribution of emitted kinetic energy directly comparable to experiments (bottom of figure \ref{fig:clus2}).
It is observed that a direct application of the rules already leads to a very good agreement with data \cite{Ber73}.  
  
However the simplified
three steps scenario described above
allows only for qualitative comparisons with data.  
In order to provide quantitative comparisons, the following
effects should be incorporated: the influence of the impact parameter and the related geometric effects,
the in-medium nucleon-nucleon collisions which distort the Fermi motion of nucleons.  
 It is worth to notice that the cluster size and multiplicity are not fixed
a priori, therefore a specific method should be developed to describe the nucleosynthesis (i.e the aggregation of nucleons) at the
early stage of the reaction. Finally, the possible secondary decay should be accounted for. A guideline for
practical implementation of these different aspects is given in ref. \cite{Lac05-2}. 
Two models (called n-IPSE\footnote{n-IPSE: nucleon-Ion Phase-Space Exploration}
\cite{Lac05}  
and HIPSE\footnote{HIPSE: Heavy-Ion
Phase-Space Exploration} \cite{Lac04}
based on the very same assumptions have been developped with a very good agreement with a large body of experimental data.
For instance, in nucleus-nucleus reactions,
charge distributions, average kinetic energies as well as the fluctuations of kinetic energies,
angular distribution, parallel velocities, event by event internal correlations, isospin diffusion effects. have
been properly reproduced in Xe+Sn reactions for beam energy from $25$ to $80$ MeV/A.
In the models, few parameters depending only on beam energy  have been adjusted
by comparison with data. These parameters are not changed when the target and/or the projectile
are changed. This aspect is essential to insure the predictivity of the models.
  
\section{Phase-space exploration: two illustrations}

A detailed comparison of the models with the experimental data
can be found in \cite{Lac04} for nucleus-nucleus and \cite{Lac05} for nucleon-nucleus reactions.
Besides the fact that comparisons are made on the phase-space after the secondary decay,
HIPSE and nIPSE models give also access to configurations before
secondary emission. This intermediate stage is of particular interest for making the link with statistical
physics. In the following, we give two examples where the models provide important information in the nuclear
context.

\subsection{Loss of memory of the entrance channel  in central events}

\begin{figure}[tbph]
\begin{center}
\includegraphics[height=8.cm,angle=-90.]{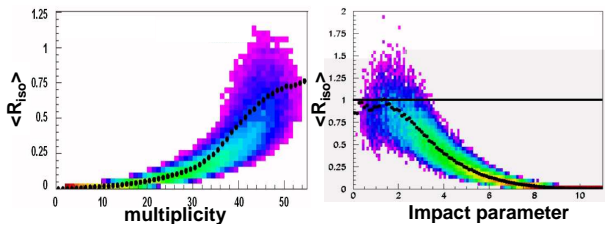}
\end{center}
\caption{Evolution of the isotropic ratio as a function of the total multiplicity (left) and  impact parameter (right).
The black circles indicate the average evolution.}
\label{fig:riso}
\end{figure}

A possible way to trace back the complete loss of memory of the entrance channel
is to consider the degree of shape equilibration. Following ref. \cite{Esc05}, we define
the isotropic ratio as $\left< R_{iso} \right> = E_\bot / 2 E_{//}$ where $ E_\bot$ and $E_{//}$ correspond respectively
to the average total transverse and parallel energy. For a spherical symmetric partition, we do expect
$\left< R_{iso} \right> = 1$, this limit corresponding to a complete shape equilibration. In figure (\ref{fig:riso}),
we have plotted the evolution of $\left< R_{iso} \right>$ as a function of multiplicity for the reaction Xe+Sn at 50 MeV/A. 
We observe an increase
of this quantity and a saturation around a value of $0.75$. A similar behavior has been found in experiments \cite{Esc05}
and assigned as a incomplete loss of the entrance channel memory. In that case, the multiplicity has been assumed
to be strongly correlated with the centrality of the reaction. In HIPSE, the average multiplicity indeed
increases with the centrality. However, large fluctuations around the mean value are also observed.
As a consequence, a given multiplicity corresponds to a large set of impact parameters (denoted by $b$).
The great advantage of HIPSE is that it gives direct access to $b$. In the left side of figure (\ref{fig:riso}), the
evolution of $\left< R_{iso} \right>$ is presented as a function of $b$. We clearly observed a saturation
around $1$ for smallest impact parameters. This observation leads to two conclusions for the model.
First, the multiplicity is not a perfect selector
of the centrality and the saturation below $1$ is a spurious saturation due to impact parameter mixing.
Second, most central events in HIPSE are completely equilibrated in shapes and a complete loss of memory of the
entrance channel is observed.
      
\subsection{Isospin equilibration}

With new radioactive beams facilities, we do expect to explore a large set of nuclei in the nuclear
chart. In HIPSE, a particular care has been taken to treat consistently the isospin degrees of freedom. For the moment,
HIPSE has only been applied to reactions with stable nuclei but can similarly be used for radioactive beams. This aspect
is illustrated in the reaction $^{48}$Ca+$^{40}$Ca at beam
energy $E_B=25$ MeV/A. In figure (\ref{fig:iso}), the $(N,Z)$ distributions before the decay
respectively for the quasi-projectile (QP) and the quasi-target (QT)
in the left side and for the mid-rapidity clusters and compound nucleus in the right side are
shown. This figure clearly shows the partial isospin equilibration between the QP and QT, while clusters in the
mid-rapidity region or close to compound nucleus are equilibrated in average. It also illustrates the
large fluctuations in the nuclear chart.
\begin{widetext}

\begin{center}
\begin{figure}[tbph]
\begin{center}
\includegraphics[height=13.cm,angle=-90.]{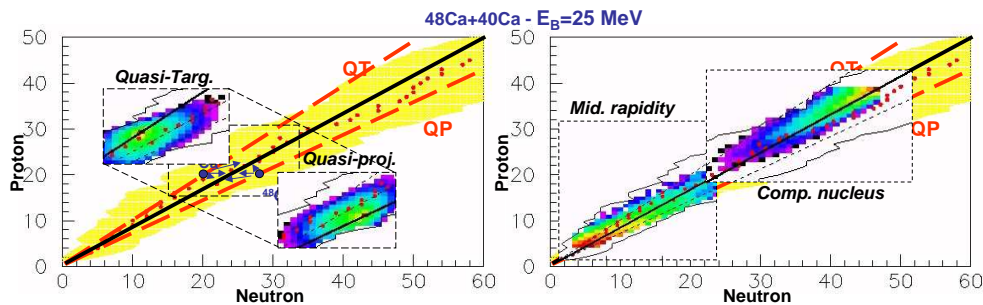}
\end{center}
\caption{Illustration of the Neutron-Proton phase-space explored for the Quasi-Target (QT), Quasi-Projectile (QP)
in the left side and for the Mid-rapidity clusters and Compound nucleus in the right side for the reaction
$^{48}$Ca+$^{40}$Ca at beam energy $E_B=25$ MeV/A (all impact parameters are mixed). In both cases, the
largest area indicates the total phase-space available (in the mass table used in HIPSE), points
indicates nuclei in the valley of stability, dashed lines correspond to the Target and Projectile $N/Z$ and
the central solid line to the equilibrium $N/Z$. }
\label{fig:iso}
\end{figure}
\end{center}
\end{widetext}
%\begin{figure}[tbph]
%\begin{center}
%\includegraphics[height=10.cm,angle=-90.]{fig_iso.eps}
%\end{center}
%\caption{Left: Evolution of the $N/Z$ of the quasi-projectile in the reaction $^{86}$Kr+$^{124}$Sn (black circles) and
%$^{86}$Kr+$^{112}$Sn (open circles) as a function of the excitation energy of the quasi-projectile. Solid line indicates the reference
%value of $^{86}$Kr while
%Right: }
%\label{fig:iso}
%\end{figure}

\section{Conclusion}
We propose simple prescriptions
that may be used to describe the pre-equilibrium emission
of clusters in the course of nuclear reactions. These rules are based on
a random sampling of the nucleons taking into account the Fermi motion
and a proper account of nuclear effects
as well as the conservation laws. Using these rules, two models dedicated respectively to
nucleon nucleus (nIPSE) and nucleus-nucleus reactions (HIPSE) around the Fermi energy have been 
developed 

Besides the good reproduction of a large variety of experimental observations on a large
set of different reactions and energies, these models underlines the importance of phase-space
explored before and after the secondary decay. This aspect is illustrated in two examples of present interest,
namely the memory loss in most central collisions and the $(N,Z)$ partitions after the
pre-equilibrium stage.

%{\bf Acknowledgments}
%We thank warmly the INDRA collaboration for permission to
%use its  data. We would like to thank V. Blideanu, O. Lopez, A. Van Lauwe and
%E. Vient for their collaboration in this work.

% For Figures insertion uncomment the next section

%\begin{figure}
%\includegraphics{figurename}
%\caption{Your caption here}
%\label{fig01} % optional figure label, must be unique
%\end{figure}

%%%%%%%%%%%%%%%%%%%%%%%%%%%%%%%%%%%%%%%%%%%%%%%%%%%%%%%%
% End of the paper
%
\end{document}